\documentclass[11pt,A4paper]{article}

\usepackage[centertags]{amsmath}
\usepackage{amssymb}
\usepackage{amsthm}
\usepackage{graphicx}
\usepackage{natbib}
\usepackage{multirow}

\renewcommand{\itemsep}{\skip0}
\renewcommand{\parsep}{\skip0}

\newcommand{\nothing}[1]{}

\newcommand{\beq}{\begin{equation}}
\newcommand{\eeq}{\end{equation}}
\newcommand{\bd}{\begin{displaymath}}
\newcommand{\ed}{\end{displaymath}}

\def\b#1{\mbox{\boldmath $#1$}}    
\def\m#1{\mbox{#1}}                

\def\ti#1{\mbox{$\tilde{\b #1}$}}

\begin{document}

\title{A hierarchical latent class model for predicting disability small area counts from survey data}

\author{Enrico Fabrizi\thanks{DISES, Universit\`a Cattolica del S. Cuore, Piacenza, Italy} \and Giorgio E. Montanari\thanks{DEFS, 
Universit\`a degli Studi di Perugia, 
Italy} \and M. Giovanna Ranalli\thanks{DEFS, Universit\`a degli Studi di Perugia, 
Italy, \texttt{giovanna.ranalli@stat.unipg.it}}}

\maketitle

\begin{abstract} \linespread{1}
\noindent This article considers the estimation of the number of severely disabled people using data from the Italian survey on Health Conditions and Appeal to Medicare. Disability is indirectly measured using a set of categorical items, which survey a set of functions concerning the ability of a person to accomplish everyday tasks. Latent Class Models can be employed to classify the population according to different levels of a latent variable connected with disability. The survey, however, is designed to provide reliable estimates at the level of Administrative Regions (NUTS2 level), while local authorities are interested in quantifying the amount of population that belongs to each latent class at a sub-regional level. Therefore, small area estimation techniques should be used. The challenge of the present application is that the variable of interest is not observed. Adopting a full Bayesian approach, we base small area estimation on a Latent Class model in which the probability of belonging to each latent class changes with covariates and the influence of age is learnt from the data using penalized splines. Deimmler-Reisch bases are shown to improve speed and mixing of MCMC chains used to simulate posteriors.

\end{abstract} 


\section{Introduction}
\linespread{1.2}

Aging of the population, in particular when it is connected with a status of disability, is a central issue for policy makers in most OECD countries. Severe disability may cause a condition of autonomy deprivation so that the individual needs personal assistance to complete basic everyday tasks. Decision makers, responsible for health organization and planning are very interested in classifying the population and monitoring the number of people with different levels of disability in order to determine the care needs in a territory and configure appropriate social policies. 

Assessment of levels of functioning and disability can be conducted on a personal basis via interviews and medical investigation by a medical staff from Local Health Departments on each person who requests benefits and extra sanitary assistance. However, such information does not provide sufficient data for classifying the population according to the different levels of disability, since it does not include all those people who do not formally inquire for extra public sanitary assistance and, also, different scales are used in the classification by each Department. It is therefore very important to be able to estimate the impact of the phenomenon using an alternative data source. Since ah hoc surveys are too expensive, we propose to use data from some existing extensive surveys on the population routinely run by National Statistical Institutes. 

In this work we look in particular at the case of Italy -- that shows the largest proportion of population aged 65 or more among EU25 European countries, $20.3\%$ in 2011. The most structured and complete  data source on disability in Italy is the Health Conditions and Appeal to Medicare Survey (HCAMS) conducted by the National Institute of Statistics every five years. The most recent available edition is that from the years 2004-05. The HCMAS employs a questionnaire that accounts for the International Classification of Impairments, Disabilities and Handicaps (ICIDH) developed by the World Health Organization in 1980 and evaluates disability by means of 14 items that include the Activities of Daily Living \citep[ADL;][]{Katz:1963}. In this survey a person is considered as disabled if, excluding temporary limitations, he/she expresses the largest degree of difficulty in at least one of a set of functions concerning the ability of a person to accomplish everyday tasks such as getting washed and dressed, eating, walking or hearing, even with the aid of tools such as glasses, walking sticks, prostheses. More details on the items are provided in Section \ref{sec:data}. From a policy maker perspective, this definition of disability is way too loose, in that a person may be defined as disabled because he/she cannot hear a TV-show, but he/she can perfectly take care of him/herself without an extra-burden for the Health Care system. In fact, what is mainly interesting for policy making and planning is to evaluate that level of  disability that is  connected with a status of dependency. 

ADLs have been used extensively to detect disability. The basic approach to measuring disability is by using a summed index, where individual scores on all items are summed to produce a total. One obvious sufficient condition for the summed index (total score) to be valid is that all items equally contribute to some sort of measure of disability, a condition that may not hold. Tools from Item Response Theory may be applied to obtain a proper aggregate measure of functional disability. \citet{erosh:2002} provides a review of application of such models to ADLs. See also \citet{irtadl} for the use of ADLs to obtain an aggregate measure of functional disability from a national survey in Spain. To classify the population according to different levels of disability, Latent Class Models \citep{Laza:1968} can be employed. \citet{Mont:Rana:Euse:2011} use this approach on data from the HCAMS to classify people and are able to identify, among the others, one class as composed by those in a condition of severe disability. In this work we follow this approach, but we extend it to tackle the following issues. 

The survey uses a stratified two stage sampling design and provides direct estimates reliable up to the Administrative Region level (NUTS2). In this work we focus, within Italy, on the case of the Administrative Region \emph{Umbria}. It is located in the center of Italy and shows the second largest proportion of people aged 65 or more in the nation ($23.1\%$). Like all Regions in Italy, Umbria is divided into 12 Health Districts and the local Administrative Department responsible for health organization and planning is interested in classifying the population according to different levels of disability for each Health District and age class (50-64; 65-74; 75 and more). Once a unit in the sample is labelled with his/her most probable latent class, reliable direct estimates of the amount of population within each District and age class cannot be obtained due to the lack of a sufficient number of observations. Small area estimation techniques  should then be used to obtain estimates for the 12 Health Districts by the 3 age classes. The basic idea of small area estimation is to introduce a model to exploit the relationship between the variable of interest and some auxiliary variables for which information at the population level is available (such information may range from population counts at the small area level to individual population records). A great deal of research has been devoted to this field of survey statistics in recent years \citep[for reviews see][]{Rao:2003,Jian:Lahi:sae:2006}. However, the challenge of the present application is that the variable of interest is not observed; it is a latent variable hidden in the items surveyed. Small area estimation techniques have not been developed for estimating the total or the mean of a latent variable. 

A first solution may be a two step procedure in which $(1)$ item responses of every unit are summarized to obtain a measurement of the latent variable; this is done by fitting a latent class model to the data and label each unit with his/her most probable class. Then $(2)$ measurements (latent class memberships) are used as a known dependent variable in a small area model \citep[e.g. a multinomial mixed effects model as in][]{ghosh:jasa:98}. This solution is unsatisfactory, since when using latent variable estimates in a regression model, the association between the real value of the latent variable and the covariates is underestimated \citep{mesbah:2004}. In addition, errors are propagated from step (1) to step (2) without any control. We propose to tackle the problem of classifying the population and getting small area estimates as a whole within a Hierarchical Bayesian framework in which the probability of belonging to each latent class changes with covariates. Age by sex by marital status counts are available for each municipality from administrative registers and can be used as covariates. A random effect capturing the variability between the districts not accounted for by the covariates is also introduced.

No parametric assumption on the functional form of the influence of age is made. We therefore extend the latent class mixed effects model with covariates to the case in which the effect of a continuous variable is only assumed to be a smooth function. Penalized spline (p-spline) regression is then used to approximate this function \citep[see][for a  general treatment of p-splines and their applications]{Eile:Marx:repl:1996,Rupp:Wan:Carr:03}. \citet{Opso:Clae:Rana:Kaue:Brei:non-:2008} exploit the mixed models representation of p-splines \citep[see][Chapt. 4]{Rupp:Wan:Carr:03} to incorporate it in a small area model. The smooth function can be modeled using natural cubic splines, B-splines, truncated polynomials, radial splines and others. In Bayesian analysis based on MCMC methods, the particular choice of basis is crucial for its consequences on the mixing  of the MCMC chains. \citet{Crai:Rupp:Wand:bugs:2005} provide an overview of the implementation of nonparametric Bayesian analysis via p-splines and advocate the use of low-rank thin-plate splines for their good mixing properties of the MCMC chains. Since we are modeling a latent variable, even the use of this basis yields a large posterior correlation for the fixed and the random coefficients in the mixed model representation of the p-spline. In this work we propose the use of the Demmler-Reinsch  basis \citep{nych:cumm:96}. Like B-splines, they provide a transformation of basis that implies an orthogonal design matrix.  The latter property allows for a much better-behaved chains and a reduction of computational time. 

The paper is organized as follows. Section \ref{sec:data} provides an overview of the survey data and the items employed to measure disability. Section \ref{sec:mod} illustrates the alternative models employed, describes the model and the p-splines specification. Section \ref{sec:results} shows the final results after the procedure of model selection and checking. Section \ref{sec:concl} provides some concluding remarks and directions for future research.

\section{The data} \label{sec:data}

The target population for the HCAMS is that of households. The survey design uses a complex sampling scheme and, in particular, it is as follows. Within a given Province (LAU1),  municipalities are classified as Self-Representing Areas (SRAs) - consisting of the larger municipalities - and Non Self-Representing Areas (NSRAs) - consisting of the smaller ones. In SRAs each municipality is a single stratum and households are selected by means of systematic sampling. In NSRAs the sample is based on a stratified two stage sample design. Municipalities are  primary sampling units (PSUs), while  households are  Secondary Sampling Units (SSUs). PSUs are divided into strata of approximately the same dimension in terms of population. One PSU is drawn from each stratum with probability proportional to the PSU population size. The SSUs are selected by means of systematic sampling in each PSU. All members of each sample household, in both SRAs and NSRAs, are interviewed. The survey is designed to provide reliable estimates at a regional level (NUTS2). Note that all those who live permanently in a care facility are not included in the sample. 

The 2004-05 edition of the HCAMS has involved  in Umbria 1,210 households and 3,088 people  divided into 40 municipalities. Since in this work we are interested in people aged 50 or more, the available sample size reduces to 1,340. In addition, Umbria is divided into 12 Health districts, that are groups of neighboring municipalities. Table \ref{tab:sampdim} provides the sample size for each subpopulation given by the intersection of each health district with each of 3 age classes. Many domains have very few observations (less than 30) available, making it very unreliable to use direct estimation techniques. 

\begin{table}
\caption{Number of observations available for each domain of interest given by Health district by age class. Health districts are ordered by code.}\label{tab:sampdim}
	\centering
		\begin{tabular}[]{l|rrr}
&	$50-64$ & $65-79$&	 $\geq 80$\\
\hline \hline
$[11]$ Alto Tevere &		32&	41&	10\\
$[12]$ Alto Chiascio&		52&	40&	18\\
\hline
$[21]$ Perugino&	117&	81&	26\\
$[22]$ Assisano&		27&	24&	8\\
$[23]$ Medio Tevere& 28&	27&	13\\
$[24]$ Trasimeno&		26&	22&	7\\
\hline
$[31]$ Valnerina&	14&	13&	5\\
$[32]$ Spoleto&		41&	45&	22\\
$[33]$ Foligno&		88&	82&	31\\
\hline
$[41]$ Terni&		120&	78&	30\\
$[42]$ Narni&		36&	35&	11\\
$[43]$ Orvieto&		39&	40&	11
\end{tabular}
\end{table}

To measure disability, the HCAMS accounts for the International Classification of Impairments, Disabilities and Handicaps (ICIDH) developed by the World Health Organization in 1980. It provides a conceptual framework for disability, which is described in three dimensions: impairment, disability and handicap. In the context of health experience an impairment is any loss or abnormality of psychological, physiological or anatomical structure or function. On the other side, a disability is any restriction or lack (resulting from an impairment) of ability to perform an activity in the manner or within the range considered normal for a human being, whilst a handicap is a disadvantage for a given individual, resulting from an impairment or a disability, that limits or prevents the fulfillment of a role that is normal (depending on age, sex, and social and cultural factors) for that individual. As we said before, in the HCAMS, a person is considered disabled if, excluding temporary limitations, he/she expresses the largest degree of difficulty in at least one of a set of functions concerning the ability of a person to accomplish everyday tasks, even with the aid of tools.  

Such functions are evaluated by means of 14 items in the questionnaire that include the Activities of Daily Living \citep[ADL;][]{Katz:1963}. Four types of disability are defined according to the kind of deprived functional autonomy: confinement, difficulties in movement, difficulties in everyday activities and tasks, sensory deprivation. A condition of permanent constriction in bed, on a chair or at one's home due to physical or psychical reasons is intended for  confinement. People with difficulties in movements show problems in walking, i.e. they can only walk few steps before taking a rest, they cannot climb the stairs without stopping, they cannot bend to pick up something from the ground. Difficulties in the activities of daily living are concerned essentially with a lack of independence in accomplishing basic everyday tasks as going to bed, sitting, getting dressed or washed, taking a bath or a shower. Finally, sensory deprivation includes limitations in hearing -- e.g. not being able to listen to a TV show even at a high volume,  in spite of the use of hearing aid; limitations in seeing -- e.g. not being able to recognize a friend at a meter distance; limitations in talking. The items are all ordinal with categories increasing with the difficulty of fulfilling the task and they are grouped according to the  aforementioned four types of disability. The items (and corresponding categorization) follow. 

\begin{itemize}
\item[1.] Confinement: 
	\begin{itemize}
	\item CONF = type of confinement (four categories: not confined, confined to one's home, confined to a chair, confined to one's bed).
	\end{itemize}
\item[2.] Difficulties in movements
	\begin{itemize}
	\item DIST = longest walkable distance (three categories: more than 200 m., less than 200 m., only a few steps);
	\item STAIR = going up and down the stairs (four categories: capable, with some effort, only with a lot of effort, not capable);
	\item STOOP = stooping down (four categories as for STAIR).
	\end{itemize}
\item[3.] Difficulties in everyday activities and tasks
	\begin{itemize}
	\item BED = getting in and out of bed (three categories: with no effort, with some effort, only with the help of somebody);
	\item CHAIR = sitting and standing (three categories as for BED);
	\item DRESS = getting dressed and undressed (three categories as for BED);
	\item BATH = taking a bath or a shower (three categories as for BED);
	\item WASH = washing one's face and hands (three categories as for BED);
	\item EAT = eating cutting one's food (three categories as for BED);
	\item CHEW = chewing (four categories as for STAIR);
	\end{itemize}
\item[4.] Sensory deprivation
	\begin{itemize}
	\item HEAR = hearing a TV show (three categories: capable, only at high volume, not capable);
	\item SIGHT = seeing and recognizing a friend (three categories: capable, only at a meter distance, not capable);
	\item SPEE = speaking (three categories as for BED).
\end{itemize}
\end{itemize}

\citet{Mont:Rana:Euse:2011} consider the same set of items to provide a classification of the elderly. In addition, they use the tools developed for the validation of Rasch models and perform item fit analysis to assess which items provide a unidimensional latent construct. Item fit analysis has been conducted on the dataset at hand here and results are in line with those found in \citet{Mont:Rana:Euse:2011}. In particular, out of the aforementioned 14 items, nine are selected that provide a valid and reliable scale. In particular, CONF, CHEW, HEAR, SEE and SPEE are removed from the analysis. This is in line with the literature that raises concerns when merging ``motor'' and ``cognitive'' items. In addition, note that CONF is indeed a so called ``summary item'' and is found to be redundant with respect to motor items, while sensory deprivation items are found to have a very little discrimination power when used in a latent class model. For all these reasons, in the following sections we employ this reduced set of nine items to fit the latent class small area model.

\section{The latent class small area model} \label{sec:mod}

In this section the proposed methodology to tackle the aforementioned problem is presented in detail. First, the structure of the model is provided: a latent class model is coupled with a small area unit level model in which the probability of belonging to each latent class is modeled via a multinomial mixed effects model. Sex, marital status and age are considered as covariates. Note that the choice for covariates in a small area model is limited to those for which population level information is available. In the present application, in fact, better predictors of disability could be the presence of morbidities or chronic diseases. Such information, although present in the HCAMS, is not supported by the knowledge of population counts in municipalities or health districts. The effect of age  does not have a pre-specified parametric form and is learnt from the data using penalized splines. Section \ref{sec:spl} provides details on this issue. Then, parameter estimates for the model are obtained within a Hierarchical Bayesian framework. The prior specification of the model is then provided, together with the form of the small area estimates. 

\subsection{Model specification} \label{sec:modsp}
Let $Y_{ijt}$ denote the response of unit $i$ within small area $j=1,\ldots,J$ on item $t=1,\ldots,T$. The number of small areas obtained by cross-classifying districts and age classes is denoted by $J=12\times 3$, the number of units in the sample for small area $j$ is $n_j$ so that the overall sample size is given by $n=\sum_{j=1}^J n_j$. The total number of items is $T=9$, while a particular level of item $t$ is denoted by $h_t$ and its number of categories by $H_t$. The latent class variable is denoted by $Q_{ij}$, a particular latent class by $c$ and the number of latent classes by $C$. The full vector of responses of unit $i$ in small area $j$ is denoted by $\b{Y}_{ij}$, whilst $\b{h}$ refers to a possible answer pattern. 

If $N_j$ is the population size of small area $j$, we are interested in estimating the small area totals $$Q_{j}(c)=\sum_{i=1}^{N_j}I(Q_{ij}=c),$$ for $j=1,\ldots,J$ and $c=1,\ldots,C$, where $I(\cdot)$ denotes the indicator function. The latent class small area model can be expressed as
\begin{equation}\label{eq:modmu}
\left\{
\begin{array}{lll}
P(\b{Y}_{ij}=\b{h}) & =  &  \sum_{c=1}^C P(Q_{ij}=c)P(\b{Y}_{ij}=\b{h}|Q_{ij}=c),\\
\displaystyle  \log \frac{P(Q_{ij} =  c)}{P(Q_{ij} =  1)} & = & \alpha_{0c}+v_{d(i)c}+ \alpha_{1c}\texttt{sex}_{ij}+f_c(\texttt{age}_{ij}) + \\ 
& & +\b{\gamma}_{c}\texttt{marital-status}_{ij}, \m{ for $c=2,\ldots,C$.}
\end{array}
\right.
\end{equation}
The first equation is the latent class model in which the probability of observing a response pattern $\b{h}$ is a weighted average of class-specific probabilities. In fact, the term $P(\b{Y}_{ij}=\b{h}|Q_{ij}=c)$ is the conditional response probability of observing pattern $\b h$ given that unit $i$ in small area $j$ belongs to class $c$, and the weight is the probability that such unit belongs to the latent class $c$. By assuming independence of responses within latent classes, the conditional probability takes the form 
\begin{equation}\label{indep:assumption}
P(\b{Y}_{ij}=\b{h}|Q_{ij}=c)= \prod_{t=1}^T P({Y}_{ijt}=h_t|Q_{ij}=c).
\end{equation} 

The probabilities $P(Q_{ij}=c)$ are modeled via the multinomial logistic mixed model of the second equation in which the first latent class is the reference one. Sex (reference category is `female') and marital status (with three binary variables -- `married', `separated/divorced', `widow' -- reference category is `single') enter this model parametrically, while  $f_c(\cdot)$, for $c=2,\ldots,C$, is a smooth function of age to be estimated nonparametrically. Details on this latter aspect of the model are provided in Section \ref{sec:spl}. Finally, $v_{d(i)c}$ is a zero mean random variable, for $d=1,\ldots,12$ and $c=2,\ldots,C$, and $d(i)$ denotes the health district which unit $i$ belongs to. These terms represent random intercepts included  for each health district to capture the between-district variation not explained by the covariates. It is common practice in small area estimation to include a random intercept to model the intra-area correlation and the area effect not accounted for the covariates in the model. 

Once prior distributions are assumed on model parameters, a Markov Chain Monte Carlo algorithm can be used to obtain $M$ samples from the joint posterior distribution. Prior choice and MCMC specifications are described in Section \ref{sec:prio}. 

Since the small area part of model (\ref{eq:modmu}) is a multinomial mixed effect model, $C-1$ different sets of fixed coefficients for sex and marital-status, of random intercepts for the district effect and of  smooth curves for age are estimated, one for each latent class but the reference one. The latent variable can be considered univariate for the item fit analysis described in Section \ref{sec:data}; therefore, a more parsimonious model can be considered that assumes ordered classes and fits a common effect for the covariates and a different intercept for each class. If we consider a  proportional odds model  \citep[see e.g.][Chapt. 7]{Agre:cate:2002}, the second equation of (\ref{eq:modmu}) becomes 
\begin{eqnarray}\label{eq:modord}
\nonumber \m{logit} P(Q_{ij}\leq c) & = & \log \frac{P(Q_{ij}\leq c)}{P(Q_{ij}> c)} \\ 
 & = &\alpha_{0c}+v_{d(i)}+ \alpha_{1}\texttt{sex}_{ij}+f(\texttt{age}_{ij}) +\b{\gamma}\texttt{marital-status}_{ij},
\end{eqnarray}
for $c=1,\ldots,C-1$. Note that in equation (\ref{eq:modord})  a common effect from the covariates for different $c$ is considered, and this is reasonable when a continuous variable is assumed to underlie the variable $Q$. On the other hand, a different intercept $\alpha_{0c}$ is included for each latent class. These values are increasing in $c$, since $P(Q_{ij}\leq c)$ increases in $c$ for a fixed value of the covariates, and the logit is an increasing function of this probability. Moreover, they can be interpreted as `cutpoints' for the continuous underlying variable, that is $Q$ is equal to $c$ when the underlying variable takes value in the interval $(\alpha_{0c-1},\alpha_{0c}]$.

\subsection{Penalized spline regression} \label{sec:spl}

The effect of age in both models (\ref{eq:modmu}) and (\ref{eq:modord}) does not have a pre-specified functional form, but is only assumed to be a smooth function. A nonparametric model has significant advantages compared with a parametric approach when the functional form of the relationship between the variable of interest and one or more covariates cannot be specified \emph{a priori}, since erroneous specification of the model can result in biased estimates. This is particularly relevant in the present application, where the variable of interest is latent and no clear way exists of predetermining a reasonable specific functional form for the effect of age. 

Despite the numerous extensions to the basic small area models present in literature, the inclusion of a nonparametrically specified term in a small area model has only  recently been considered, due to the methodological difficulties of incorporating smoothing techniques into the estimation tools used to this end. \citet{Opso:Clae:Rana:Kaue:Brei:non-:2008} exploit the close connection between penalized splines (p-splines) and linear mixed models \citep[see e.g.][Chapt. 4]{Rupp:Wan:Carr:03} to incorporate a nonparametric mean function specification into a small area unit level model. The p-spline representation of the smooth function $f_c(\texttt{age}_{ij})$ in (\ref{eq:modmu}) is given by 
\begin{equation}\label{eq:spli}
f_c(\texttt{age}_{ij};\beta_{c}, b_{1c}, \ldots, b_{Kc}) =  \beta_{c}\texttt{age}_{ij} + \sum_{k=1}^Kb_{kc}z_{ijk},
\end{equation}
where $\beta_{c},b_{1c}, \ldots, b_{Kc}$ are regression coefficients, $z_{ijk}$ are  basis functions used to model the smooth function, e.g. natural cubic splines, B-splines, truncated polynomials, radial splines and others, that depend on a set of fixed knots $\kappa_1<\kappa_{2}<\ldots<\kappa_{K}$. The number of knots $K$ has to be large enough to ensure the desired flexibility \citep[][Chapt. 5, suggest to use one knot every 4 or 5 unique values for the covariate, with a maximum number of 35]{Rupp:Wan:Carr:03}, and they should be chosen to be nicely scattered over the range of the covariate, with quantiles being a good choice. As of the choice of the spline basis, while it makes a little difference in a classical framework, it is particularly relevant in a Bayesian approach, for its consequences on the mixing properties of the MCMC chains. \citet{Crai:Rupp:Wand:bugs:2005} offer details for p-splines smoothing within a Bayesian framework and advocate the use of thin-plate splines, for which they notice a  posterior correlation of parameters that is much smaller than for other basis. In this case $z_{ijk}=|\texttt{age}_{ij}-\kappa_{k}|^{3}$. 

To avoid overfitting, the magnitude of the coefficients of the basis $b_{kc}$, for $k=1,\ldots,K$, is penalized by shrinking it towards zero. This can  be accomplished by considering $\beta_c$ as an independent parameter and  $\b b_{c}=(b_{1c}, \ldots, b_{Kc})^{T}$ as a vector of random parameters with $E(\b b_{c})=0$ and $E(\b b_{c} \b b_{c}^{T})=\sigma_{bc}^{2}\b\Omega_{K}^{-1}$, where the  $(l,k)$-th entry of $\b{\Omega}_{K}$ is given by $|\kappa_l-\kappa_k|^3$. 

In this paper we propose to use the Demmler-Reinsch orthogonalization of a set of basis functions (any set, not necessarily thin-plate splines) to improve the mixing of the MCMC chains and eliminate the posterior correlation of $\beta_c$ and of $\b b_c$. To this end, the following steps have to be implemented. In particular, given $z_{ijk}$, let $\ti{z}_{ij}=(1,\texttt{age}_{ij},z_{ij1},\ldots,z_{ijK})$, let $\ti{Z}$ be the $n\times(K+2)$ matrix whose $ij$-th row is given by $\ti{z}_{ij}$, let $\ti b_c=(\alpha_{0c},\beta_c,\b b_c^T)^T$ and let 
\begin{displaymath}
\b{D} = 
\begin{bmatrix}
\b{0}_{2\times 2} & \b 0_{2\times K} \\
\b 0_{K\times 2} & \b \Omega_{K} \\
\end{bmatrix}.
\end{displaymath}
Then, let $\ti{R}$ be a square $(K+2)\times(K+2)$ upper triangular invertible matrix so that $\ti R^T \ti R$ is the Cholesky decomposition of $\ti Z ^T \ti Z$. The singular value decomposition of the symmetric matrix $\ti R^{-T} \b D \ti R^{-1}$, where $\ti R^{-T}$ is the transpose of $\ti R^{-1}$, is given by $\ti U \m{diag}(\ti s)\ti U^T$. For ease of notation let the entries of the vector $\ti s$ be arranged in an increasing fashion. Then, given the nature of $\b D$, the first two entries of $\ti s$ are equal to zero, i.e. $\ti s =(0,0,\b s^T)^T$. 

Now, if we compute $\ti A = \ti Z \ti R^{-1} \ti U$, then the columns of $\ti A$ identify the set of basis functions known as the Demmler-Reinsch basis and show the orthogonality property. In addition, given the ordering of $\ti s$, and then of the columns of $\ti U$, the columns of $\ti A$ will exhibit more oscillations from the first to the $(K+2)$-th and, in particular, they will intersect the $x$-axis $\nu -1$ times, for $\nu=1,\ldots,K+2$ \citep[see][for more details on these aspects of Demmler-Reinsch bases]{nych:cumm:96}. 

Then, the transformation is given by 
\begin{equation}\label{eq:DRtransf}
\ti Z\ti b_c=\ti Z \ti R^{-1} \ti U \ti U^T \ti R \ti b_c=\ti A \ti w_c,
\end{equation}
where $\ti w_c=(\omega_{0c},\omega_{1c},\b w_c^T)^T$. If we denote by $\b A$ the matrix made of the last $K$ columns of $\ti A$, then  equation (\ref{eq:spli}) can be rewritten as  
\begin{equation}\label{eq:spliDEMM}
f_c(\texttt{age}_{ij};\beta_{c}, w_{1c}, \ldots, w_{Kc}) =  \beta_{c}\texttt{age}_{ij} + \sum_{k=1}^Kw_{kc}a_{ijk},
\end{equation}
where $a_{ijk}$ is the $(ij,k)$-th element of $\b A$. In (\ref{eq:spliDEMM}), $\beta_c$  is considered like  in (\ref{eq:spli}) as a fixed parameter, while $\b w_c$ as a vector of random parameters with $E(\b w_{c})=\b 0$ and $E(\b w_{c} \b w_{c}^{T})=\sigma_{bc}^{2}\m{diag}(\b s)^{-1}$. The latter form of the variance of $\b w_c$ is obtained as follows. From (\ref{eq:DRtransf}) $\b w_c = \b U^T \b R_{22} \b b_c$, where $\b U^T$ and $\b R_{22}$ are defined from 
\begin{displaymath}
\ti{U} = 
\begin{bmatrix}
\b{I}_{2\times 2} & \b 0_{2\times K} \\
\b 0_{K\times 2} & \b U \\
\end{bmatrix}\quad \m{and} \quad 
\ti{R} = 
\begin{bmatrix}
\b{R}_{11} & \b{R}_{12} \\
\b 0_{K\times 2} & \b R_{22} \\
\end{bmatrix},
\end{displaymath}
respectively. Then, $$E(\b w_{c} \b w_{c}^{T})=\sigma^2_{bc} \b U^T \b R_{22} \b\Omega_K^{-1} \b{R}_{22}^T\b U=\sigma^2_{bc} \b U^T (\b U \m{diag}(\b s)\b U ^T)^{-1} \b U=\sigma^2_{bc} \m{diag}(\b s)^{-1},$$
because 
\begin{displaymath}
\ti R^{-T} \b D \ti R^{-1} = 
\begin{bmatrix}
\b{0}_{2\times 2} & \b 0_{2\times K} \\
\b 0_{K\times 2} & \b R_{22}^{-T}\b\Omega_K\b R _{22}^{-1} \\
\end{bmatrix},
\end{displaymath}
and 
\begin{displaymath}
\ti U \m{diag}(\ti s)\ti U^T = 
\begin{bmatrix}
\b{I}_{2\times 2} & \b 0_{2\times K} \\
\b 0_{K\times 2} & \b U \m{diag}(\b s)\b U ^T \\
\end{bmatrix}.
\end{displaymath}

\subsection{Prior specification} \label{sec:prio}
To complete the Bayesian specification of the model we need to choose priors for the parameters involved in equation (\ref{eq:modmu}): the conditional probabilities $P(\b Y_{ij}=\b{h}|Q_{ij}=c)$, the regression parameters ($\alpha_{0c}$, $\alpha_{1c}$, $\beta_c$, $\b \gamma_c$), for $c=2,\ldots,C$ and the two sets of random effects $\b{v}_c$, $\b{b}_c$ -- or $\b{v}_c$, $\b{w}_c$ -- for $c=2,\ldots,C$.  For the conditional probabilities in (\ref{indep:assumption}) we assume
\[
P(Y_{ijt}=h_t|Q_{ij}=c) \sim \m{Dirichlet}(\b{1}_{H_t})
\]
for $t=1,\dots,T$, where $\b{1}_{H_t}$ is a vector of ones with a length given by $H_t$, i.e. by the number of categories of item $t$. The choice of this prior is customary and implies independent uniform priors on the probability of selecting each level of the item being considered.

For the regression parameters $\alpha_{0c}$, $\alpha_{1c}$, $\beta_c$ and  $\b \gamma_c$ we assume diffuse normal priors with mean 0 and variance $100$ that are sufficiently non-informative and computationally more convenient than flat priors over the real line.

For the random effects multivariate normality is assumed. For the health district effects, we also assume a priori independence: 
\begin{equation}
v_{dc} \stackrel{ind}{\sim} N(0,\sigma^2_{vc}), \nonumber
\end{equation}
 for $d=1,\ldots,12$ and $c=2,\dots,C$. On the other hand, for the random effects used in the p-spline representation, consistently with section \ref{sec:spl} we have: 
\begin{equation}
\b b_{c} \sim \m{MVN} (\b{0},\sigma_{bc}^{2}\b\Omega_{K}^{-1}), \nonumber
\end{equation}
when thin-plate splines are used -- equation (\ref{eq:spli}) -- and $$\b w_{c} \sim \m{MVN} (\b{0},\sigma_{bc}^{2}\m{diag}(\b s)^{-1}),$$ when the Demmler-Reinsch orthogonalization is employed -- equation (\ref{eq:spliDEMM}). The normality of the random effects is a standard assumption in hierarchical models including those used in representing splines \citep{Crai:Rupp:Wand:bugs:2005}; moreover departures from normality of random effects are very hard to detect \citep{fabri:trivi:2009}.

To complete the prior specification, we need to set prior distributions for the variance components $\sigma_{bc}^2$ and $\sigma_{vc}^2$ for $c=1,\ldots,C$. This task is critical as in Bayesian mixed models the posterior distributions of these parameters are known to be sensitive to prior specification \citep{Gelman:2006}. In particular, the choice of the prior for $\sigma_{bc}^{2}$ is crucial as this parameter influences the smoothness of the p-splines. We assume the following prior for the random effects standard deviation: 
\[\sigma_{bc} \sim \m{Uniform} (0,B)
\]
$c=1,\dots,C$. The use of uniform priors for the square root of the variance components is discussed in \citet{Gelman:2006}. The same prior is adopted for the standard deviation associated to the health district effects: 
\[
\sigma_{vc} \sim \m{Uniform} (0,B).
\]
The choice of the prior for the variance components along with a sensitivity analysis exercise and a discussion of how the value of $B$ has been selected is contained in Section \ref{check:sens}. When the small area model is based on equation (\ref{eq:modord}), the prior specification is the same for the subset of regression coefficients and random components considered. 

\subsection{Small area estimates}
The posterior distributions for the parameters of our model cannot be obtained in closed form. We use MCMC software  to generate samples from the posterior distributions. 
The posterior distributions of the counts for the different disability classes in the age by district domains are simulated combining draws from the posterior distributions of model parameters and population counts accurately known from population registers. In particular, from administrative registers, we know \texttt{age $\times$ sex $\times$ marital-status $\times$ district} population counts $N_{\ell}$, for $\ell=1,\ldots, L=50\times 2 \times 4 \times 12$. Note that 50 is the number of different ages of persons in the sample (they range from 51 to 100). Specifically, the posterior mean of the counts for the class $c$ in small area $j$ is obtained as follows:
\begin{equation}\label{est.counts}
\hat{Q}_{j}(c)=\frac{1}{M}\sum_{m=1}^{M}\sum_{\ell \in \ell_j}N_{\ell}P^{(m)}(Q_{\ell}=c|v_{d(\ell)c},\texttt{age}_{\ell},\texttt{sex}_{\ell},\texttt{marital-status}_{\ell})
\end{equation}
where 
\begin{itemize}
\item $m$ indexes the MCMC runs and $M$ is their number, 
\item $\ell$ is a particular cell and $\ell_j$ is the subset of cells from the total $L$ that is contained in the small area $j$, 
\item and the probabilities $P^{(m)}(Q_{\ell}=c|v_{d(\ell)c},\texttt{age}_{\ell},\texttt{sex}_{\ell},\texttt{marital-status}_{\ell})$ are given by 
\begin{eqnarray}
&&P^{(m)}(Q_{\ell}=c|v_{d(\ell)c},\texttt{age}_{\ell},\texttt{sex}_{\ell},\texttt{marital-status}_{\ell})= \nonumber\\
&&=\frac{\exp\big(\alpha_{0c}^{(m)}+v_{d(\ell)c}^{(m)}+ \alpha_{1c}^{(m)}\texttt{sex}_{\ell}+f_c^{(m)}(\texttt{age}_{\ell}) +\b{\gamma}^{(m)}_{c}\texttt{marital-status}_{\ell}\big)}
{1+\sum_{r=2}^{C}\exp\big(\alpha_{0r}^{(m)}+v_{d(\ell)r}^{(m)}+ \alpha_{1r}^{(m)}\texttt{sex}_{\ell}+f_r^{(m)}(\texttt{age}_{\ell}) +\b{\gamma}^{(m)}_{r}\texttt{marital-status}_{\ell}\big)},
 \nonumber
\end{eqnarray}
in the case of the multinomial logistic model, and by $$P^{(m)}(Q_{\ell}=c|v_{d(\ell)},\texttt{age}_{\ell},\texttt{sex}_{\ell},\texttt{marital-status}_{\ell})=P^{(m)}(Q_{\ell}\leq c)- P^{(m)}(Q_{\ell}\leq c-1)$$ in the case of the proportional odds model, where 
\begin{eqnarray}
&&P^{(m)}(Q_{\ell}\leq c|v_{d(\ell)},\texttt{age}_{\ell},\texttt{sex}_{\ell},\texttt{marital-status}_{\ell})= \nonumber\\ 
&&= \frac{\exp\big(\alpha_{0c}^{(m)}+v_{d(\ell)}^{(m)}+ \alpha_{1}^{(m)}\texttt{sex}_{\ell}+f^{(m)}(\texttt{age}_{\ell}) +\b{\gamma}^{(m)}\texttt{marital-status}_{\ell}\big)}
{1+\exp\big(\alpha_{0c}^{(m)}+v_{d(\ell)}^{(m)}+ \alpha_{1}^{(m)}\texttt{sex}_{\ell}+f^{(m)}(\texttt{age}_{\ell}) +\b{\gamma}^{(m)}\texttt{marital-status}_{\ell}\big)},
 \nonumber
\end{eqnarray}
and similarly for $P^{(m)}(Q_{\ell}\leq c-1)$.
\end{itemize}
Posterior standard deviations, quantiles and other summaries of the counts' posterior distribution can be obtained similarly.

\section{Results} \label{sec:results}

\subsection{Model selection}
Before proceeding with the analysis, several model choices need to be taken. Primarily,  the number of latent classes has to be chosen together with the specification of the model for $P(Q_{ij}=c)$ (i.e. multinomial logit vs proportional odds model). The Deviance Information Criterion \citep{Spiegel:etal:2002} is often used, especially in association with MCMC estimation, for model selection purposes. Nonetheless, for the latent class model we are considering, the calculation of the penalty term in the Deviance Information Criterion is notoriously problematic. To circumvent this problem, we explore the fit of alternative models using the deviance $D(\b{Q})=-2\log L(\b{Y}|\b{Q})$, where $\b{Y}$ and $\b{Q}$ are a shortcut notation for the observed data and the latent variables. Note that considering the likelihood indexed on class memberships is consistent with our aim of estimating counts related to these classes. Using MCMC draws we may calculate the posterior mean of the deviance, an approximation of the frequentist deviance, or some other summary measure of its posterior distribution. \citet{Aitkin:etal:2009} propose to consider the whole posterior Cumulative Distribution Function (CDF) of the deviance. Plotting the deviances CDF provides information not only about their location but also about the complexity of the underlying models: \citet{Aitkin:etal:2009} using an asymptotic argument, argue that deviances associated to more complex models will be characterized by CDFs with lower slopes.

\begin{figure}
\centering
\includegraphics[width=160mm]{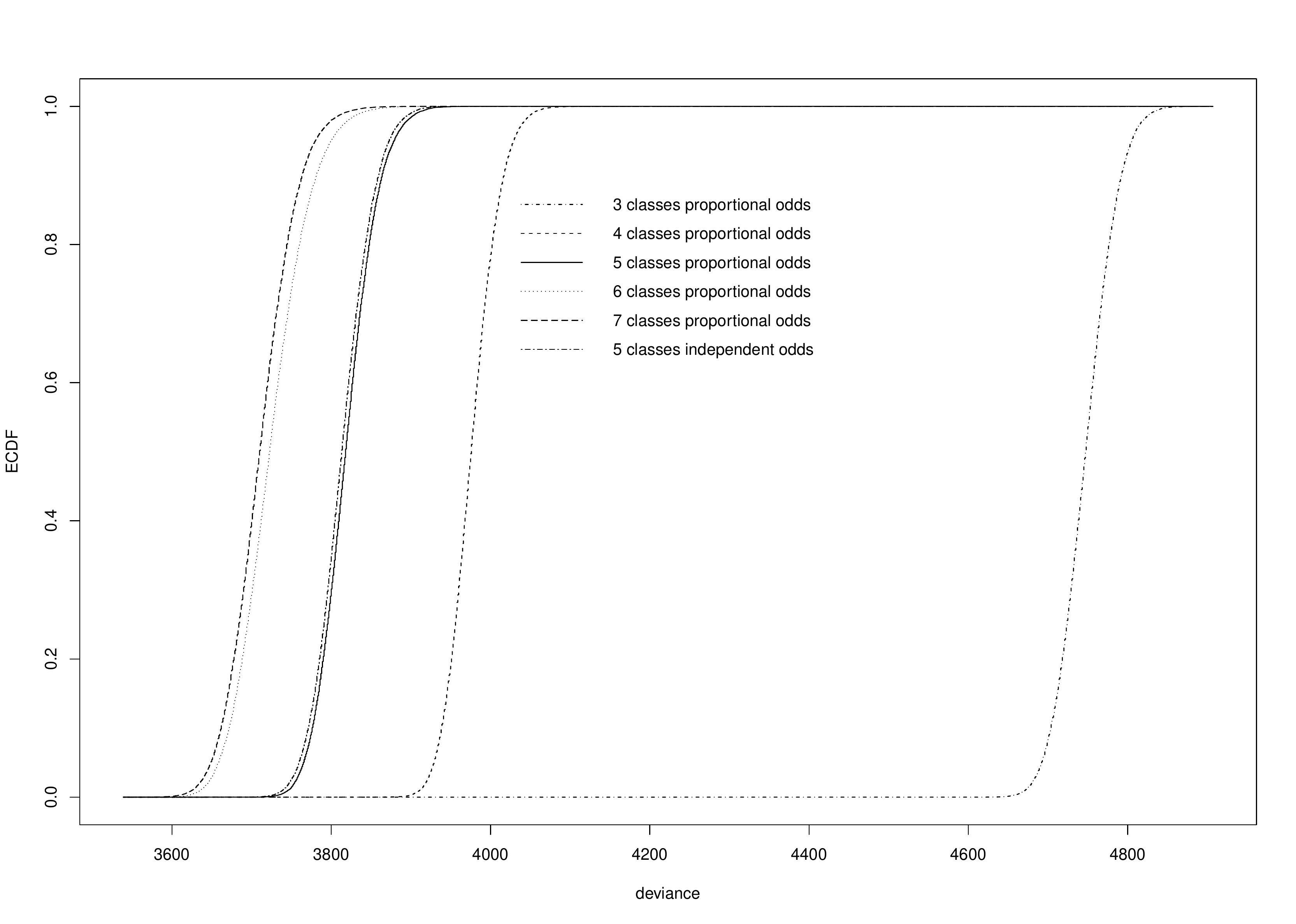}
\caption{Deviances' CDFs associated to alternative models.}\label{figure:ms}
\end{figure}

The CDFs of the alternative models we are considering are plotted in Figure \ref{figure:ms}. As the number of latent classes increases, the distribution functions move left but moves get gradually smaller. No appreciable gain is obtained when passing from 6 to 7 classes. Now, we should emphasize that the model we are proposing  is a tool to obtain reliable estimates of counts of people characterized by different  levels of disability in the age groups and health districts of Umbria; beside goodness of fit, clear interpretability of the classes in terms of response patterns for the 9 items and precision of the counts estimators should be taken into careful consideration when selecting the model. For this reason, even though Figure \ref{figure:ms} suggests the selection of a model with 6 classes, we adopt a model with 5 since it leads to classes that are more easily interpretable in terms of the predicted probabilities $P({Y}_{ijt}=h_t|Q_{ij}=c)$. In fact, when we move from the model with 5 to models with more classes, the classes at the extreme of the distribution approximately keep their composition, while those in the middle are subdivided a little confusedly. These new middle classes are not only difficult to interpret in terms of a different level of disability, but tend to be very small in terms of sample sizes, thereby leading to unstable estimates of the counts in the small areas.

The comparison between independent and proportional odds in the specification of the prior for $P(Q_{ij}=c)$ is illustrated only for the model with 5 latent classes: the two models perform very closely and similar evidences may be obtained for models with a different number of latent classes. For this reason we consider the more parsimonious proportional odds model. 

We also exclude from the chosen model the `marital status' covariate as in all models it gives no appreciable contribution to the improvement of the fit and the associated slope parameters have diffuse posterior distributions centered approximately around 0. Finally, note that the number and placement of the knots for the p-spline approximation of the effect of age has been kept fixed in all models. In particular, 12 knots have been used and chosen to be placed at the quantiles of the distribution of age (recall that we are considering only people aged 50 or more and we have 50 unique values of age).

\subsection{Model checking and sensitivity analysis}\label{check:sens}
In this section we illustrate a sensitivity analysis exercise to assess the impact of the selected prior distributions and parameters on the estimates of interest. Let's focus first on the prior for $\sigma^2_b$, a parameter that rules the smoothness of the p-spline of equation (\ref{eq:modord}). The sensitivity exercise is also useful to motivate the choice of the prior for $\sigma^2_{b}$ introduced in Section \ref{sec:prio}. Recall that we have selected a proportional odds model for the probability of belonging to a latent class, and therefore we have a single function for the effect of age and a single $\sigma^2_{b}$, instead of the $\sigma^2_{bc}$. 

\begin{figure}
\centering
\includegraphics[width=160mm]{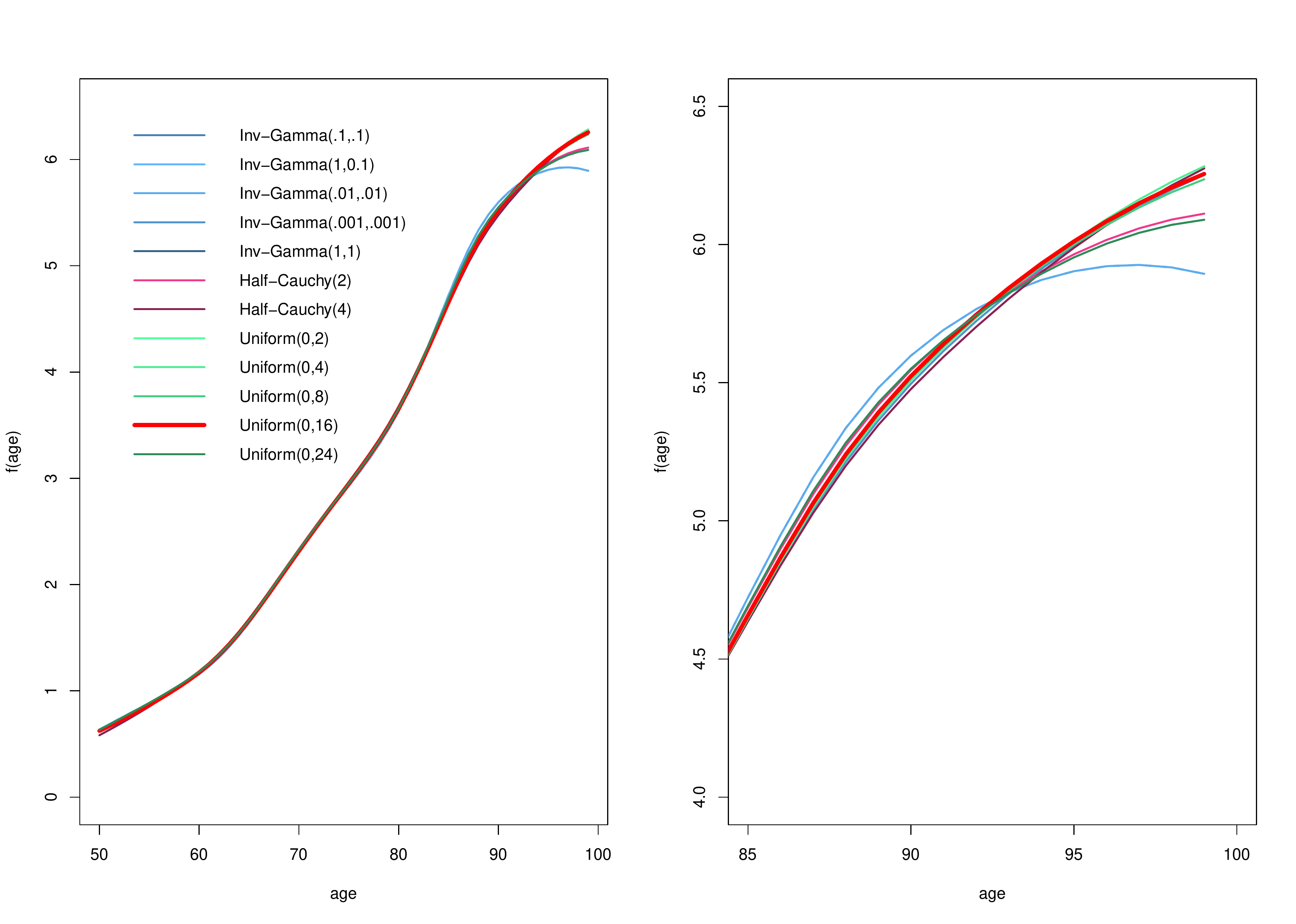}
\caption{Sensitivity of $f(\texttt{age})$ to priors on $\sigma^2_b$ }\label{figure:sens12}
\end{figure}

We consider three popular classes of priors: \textit{i}) the $\m{Inv-Gamma}(a,b)$ for $\sigma^2_{b}$; \textit{ii}) the $\m{Uniform} (0,B)$ on $\sigma_{b}$ and \textit{iii}) the $\m{Half-Cauchy}(S)$ on $\sigma_{b}$. See \citet{Gelman:2006} for motivation and details about these priors for variance components. For all the three classes of priors we consider different sets of values for their parameters. A graphical evaluation of the impact that the considered priors have on the posterior expectation of $f(\texttt{age})$ in (\ref{eq:modord}) is shown in Figure \ref{figure:sens12}. As expected, regardless of the chosen distribution, the more diffuse is the prior, the more curved is the p-spline. Nonetheless, because of the large sample size, this effect is weak except for the extreme ages (for which we have fewer observations). Also in this part of the plot most priors lead to almost identical curves and observed differences have no impact on classification of individuals with respect to disability. The only exception is represented by $\m{Inv-Gamma}(a,b)$ with `small' parameters. 

The $\m{Half-Cauchy}(S)$ priors have not been considered in estimation since they slow down the MCMC computations with respect to other available options. We decide to use the $\m{Uniform} (0,B)$ priors as they show little sensitivity to the choice of $B$, at least in the range we consider. The value $B=16$ is chosen as it is `large' with respect to the size of the effects (as suggested in the literature) but not so extreme to produce a too curved spline.

Similar sensitivity exercises have been conducted also for the variance component $\sigma^2_v$ from which it emerges that the posterior distributions of the districts effect are largely unaffected by the choice of the prior among those we have considered. The same prior adopted for $\sigma^2_b$ is then adopted for simplicity.

We use posterior predictive model checking to assess the adequacy of the estimated model. As we are working with multivariate outcome data we base our checks on the following idea, that we take from \citet{Crespi:Boscardin:2009}. If the model fits the $i$-th observation in the $j$-th area well, then $ d(\b{Y}_{ij},\b{Y}^\star_{ij}) \cong d(\b{Y}^{\star}_{ij},\b{Y}^{\star \star}_{ij})$ where $\b{Y}^{\star}_{ij}$, $\b{Y}^{\star \star}_{ij}$ are two random vectors independently drawn from the posterior predictive dstribution and $d(\b{a},\b{b})$ is some distance between T-vectors. Otherwise, if the fit is bad $ d(\b{Y}_{ij},\b{Y}^\star_{ij}) > d(\b{Y}^{\star}_{ij},\b{Y}^{\star \star}_{ij})$. We consider $d(\b{a},\b{b})=\sum_{h=1}^T \left|a_h -b_h \right|$. 

Using MCMC output we estimate posterior $p$-values, i.e. $P(d(\b{Y}_{ij},\b{Y}^\star_{ij}) > d(\b{Y}^{\star}_{ij},\b{Y}^{\star \star}_{ij})|\b{Y}),$  for each sampling unit, focusing on individuals with at least some minor problem in fulfilling daily tasks.  The reason why we have excluded the individuals who showed no problem at all in fulfilling daily tasks is technical: the class they belong to is straightforward to predict; nonetheless when we generate observations from the posterior predictive distribution, the probability that one of the answers for the nine items signals minor problems is not negligible (the first class includes also people with minor problems in fulfilling daily tasks), so the associated posterior $p$-value tend to be large although these units are properly assigned to the right class.

A summary of the estimated posterior p-values across the individuals in the sample is shown in Table \ref{table:ppv}. Only a small fraction of observations appears to be misfit.  The minimum individual p-value is $0.321$, a large and non problematic value. Only 2.5$\%$ of the sample has an associated p-value greater than $0.874$. If we look at the data we find that these sample units are, in most cases, relatively young people in a severe state of disability. As the model uses only age and gender as individual level covariates, it cannot predict the actual level of disability for these people.

\begin{table}
\begin{center}
\caption{Distribution of the estimated posterior $p-$ values across the sample}
		\begin{tabular}[]		{c | c c c c c c c}
summary& $p_{.025}$ & $p_{.25}$ & mean &  $p_{.5}$&$p_{.75}$ & $p_{.975}$\\
\hline
estimate & 0.358 & 0.478 & 0.586 & 0.594 & 0.843 & 0.874 
\end{tabular} \label{table:ppv}
\end{center}
\end{table}

\subsection{Analysis of the posterior distributions}
\begin{figure}
\centering
\includegraphics[width=160mm]{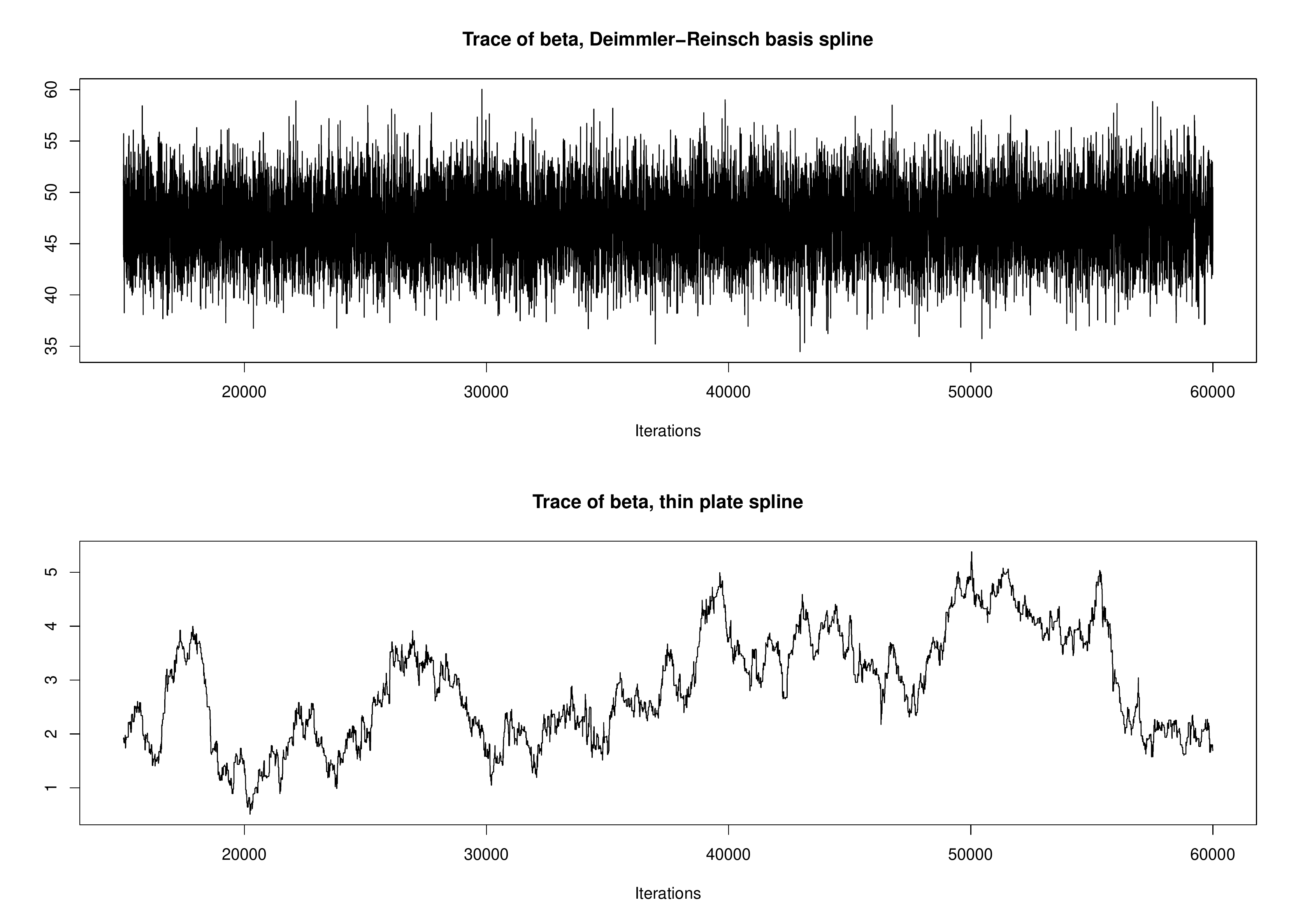}
\caption{Traces of MCMC samples from the $\beta_c$ chains for models using either Deimmler-Reinsch and thin-plate basis. Burn in omitted, thinning parameter equal to 3.}\label{figure:betatraces}
\end{figure}

To generate samples from the posterior distribution we use the OpenBugs software \citep{Openbugs:2006}. The OpenBugs code is available as Supplementary material on the web site of the Journal. We generate 120,000 MCMC runs, with a conservative burn in of 15,000 and keeping 1 in 3 replicates for analyzing the posterior. The mixing of chains for p-spline coefficients is very good, thanks to the transformation to the Demmler-Reinsch basis. With respect to thin plate splines suggested in \cite{Crai:Rupp:Wand:bugs:2005}, the mixing is dramatically improved not only for the random effects connected to the p-splines, but also for the parameters characterizing the linear component of (\ref{eq:modord}). A plot contrasting the traces of $\beta$ in the model using Deimmler-Reinsch bases with those based on the thin-plate splines can be found in Figure \ref{figure:betatraces}. The better mixing allows for shorter chains and less severe thinning, implying a relevant reduction of the computational burden.

In Table \ref{table:postprob} we report the posterior means of the conditional probabilities $P(Y_{ijt}=h_t|Q_{ij}=c)$. The five classes are easily interpretable in terms of increasing levels of difficulty in accomplishing daily tasks. While the first class is composed by individuals with essentially no problem with respect to all items, members of the second class experience problems in movements (walking long distance, going up and down the stairs) but have minor problems with respect to other items. In central classes we observe a general deterioration with respect to all items, but members are still able to wash and eat without the help of others, although with some effort. The last class is composed by those who need assistance in accomplishing most, if not all tasks of their daily life. We note that more than $75\%$ of the sample is assigned to the first class. 

\begin{table}
\begin{center}
\caption{Conditional probabilities $P(Y_{ijt}=\b{h}_t|Q_{ij}=c)$. Beside each class' label, within brackets, the percentage of sampled individuals assigned to the class}

		\begin{tabular}[]		{c | c c c c c c c c c}
	&	 \multicolumn{ 9}{c}{Class 1: without difficulties ($75.8\%$)}\\
Item level&	Bath	&	Bed	&	Chair	&	Dist	&	Dress	&	Eat	&	Stair	&	Stoop	&	Wash\\
		\hline	
1	&	0.99	&	1.00	&	1.00	&	0.99	&	1.00	&	1.00	&	0.98	&	0.97	&	1.00\\
2	&	0.01	&	0.00	&	0.00	&	0.01	&	0.00	&	0.00	&	0.02	&	0.03	&	0.00\\
3	&	0.00	&	0.00	&	0.00	&	0.00	&	0.00	&	0.00	&	0.00	&	0.00	&	0.00\\
4	&		&		&		&		&		&		&	0.00	&	0.00	&	\\
	\hline \hline
	&	 \multicolumn{ 9}{c}{Class 2: with difficulties in movements ($11.5\%$)}\\
Item level&	Bath	&	Bed	&	Chair	&	Dist	&	Dress	&	Eat	&	Stair	&	Stoop	&	Wash\\
		\hline
1	&	0.81	&	0.83	&	0.93	&	0.66	&	0.94	&	0.99	&	0.18	&	0.13	&	0.98\\
2	&	0.15	&	0.16	&	0.07	&	0.33	&	0.05	&	0.01	&	0.72	&	0.74	&	0.01\\
3	&	0.04	&	0.01	&	0.01	&	0.01	&	0.01	&	0.01	&	0.07	&	0.13	&	0.01\\
4	&		&		&		&		&		&		&	0.02	&	0.01	&	\\
	\hline \hline
	&	 \multicolumn{ 9}{c}{Class 3: with difficulties in movements and daily tasks ($6.1\%$)}\\
Item level&	Bath	&	Bed	&	Chair	&	Dist	&	Dress	&	Eat	&	Stair	&	Stoop	&	Wash\\
		\hline
1	&	0.07	&	0.19	&	0.47	&	0.53	&	0.14	&	0.86	&	0.12	&	0.10	&	0.76\\
2	&	0.58	&	0.71	&	0.52	&	0.42	&	0.75	&	0.09	&	0.50	&	0.40	&	0.22\\
3	&	0.35	&	0.10	&	0.01	&	0.06	&	0.11	&	0.05	&	0.35	&	0.39	&	0.02\\
4	&		&		&		&		&		&		&	0.03	&	0.11	&	\\
	\hline \hline
	&	 \multicolumn{ 9}{c}{Class 4: partial dependency ($1.9\%$)}\\
Item level&	Bath	&	Bed	&	Chair	&	Dist	&	Dress	&	Eat	&	Stair	&	Stoop	&	Wash\\
		\hline
1	&	0.04	&	0.07	&	0.07	&	0.09	&	0.08	&	0.56	&	0.03	&	0.03	&	0.70\\
2	&	0.25	&	0.60	&	0.79	&	0.45	&	0.59	&	0.40	&	0.07	&	0.11	&	0.25\\
3	&	0.71	&	0.33	&	0.13	&	0.46	&	0.33	&	0.04	&	0.42	&	0.29	&	0.05\\
4	&		&		&		&		&		&		&	0.48	&	0.57	&	\\
	\hline \hline
	&	 \multicolumn{ 9}{c}{Class 5: complete dependency ($4.8\%$)}\\
Item level&	Bath	&	Bed	&	Chair	&	Dist	&	Dress	&	Eat	&	Stair	&	Stoop	&	Wash\\
1	&	0.02	&	0.02	&	0.02	&	0.03	&	0.02	&	0.13	&	0.02	&	0.01	&	0.04\\
2	&	0.02	&	0.06	&	0.13	&	0.06	&	0.04	&	0.38	&	0.02	&	0.02	&	0.30\\
3	&	0.97	&	0.93	&	0.85	&	0.91	&	0.94	&	0.49	&	0.06	&	0.05	&	0.66\\
4	&		&		&		&		&		&		&	0.91	&	0.91	& \\	
\hline
\end{tabular} \label{table:postprob}
\end{center}
\end{table}

\begin{figure}
\centering
\includegraphics[width=160mm]{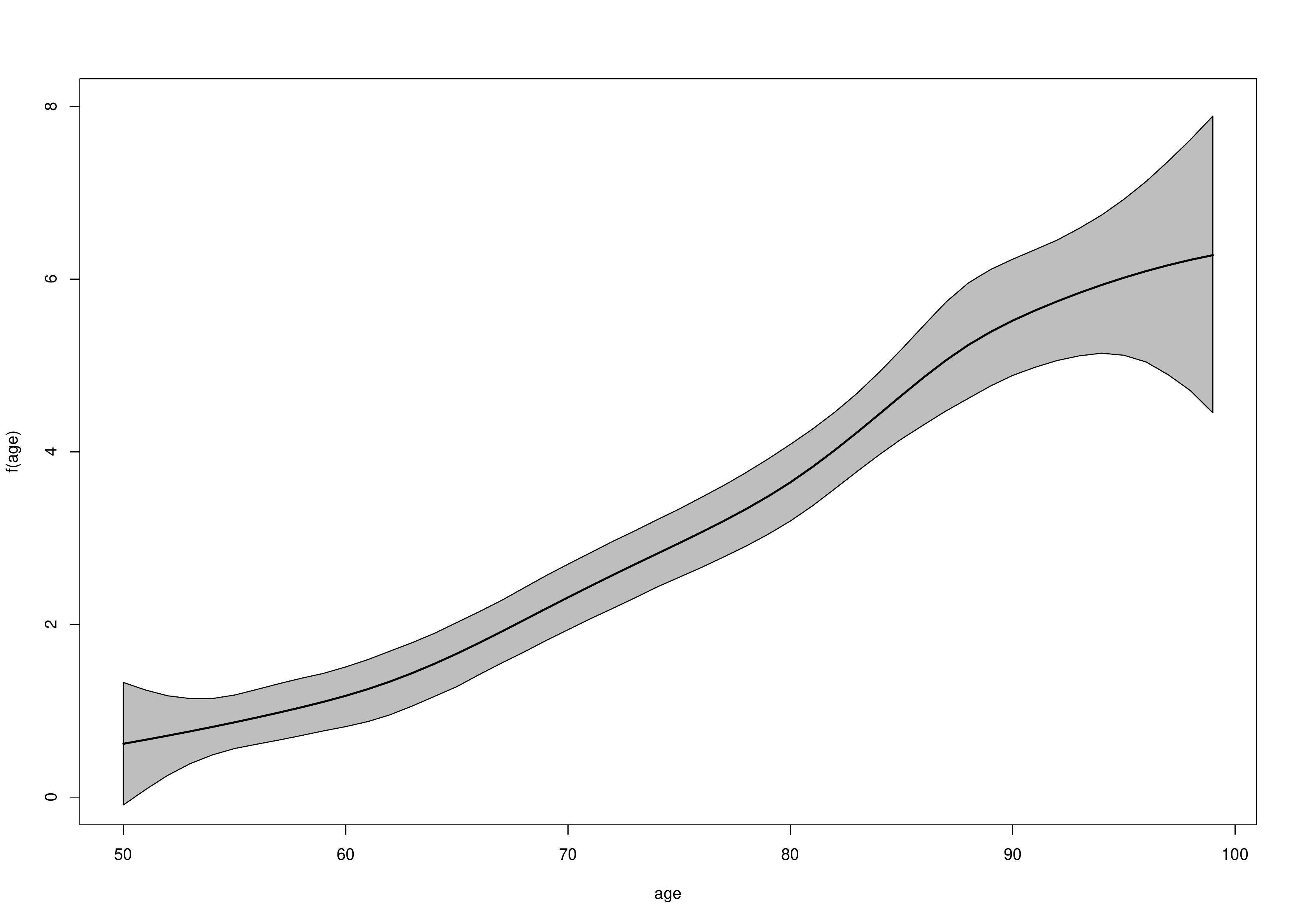}
\caption{The posterior mean$f(\texttt{age})$ (black line) with a $95\%$ probability interval (grey area) based on $0.025$ and $0.975$ posterior quantiles}\label{figure:shapeofthespline}
\end{figure}

We already illustrated the shape of $f(\texttt{age})$ when discussing its sensitivity to the prior on variance components in section \ref{check:sens}. Its posterior mean along with a $95\%$ probability interval based on posterior $0.025$ and $0.975$ quantiles is plotted in Figure \ref{figure:shapeofthespline}. The other covariate included in equation (\ref{eq:modord}) is gender. We find that the posterior mean of $\alpha_1$ is negative ($E(\alpha_1|\b{Y})=-0.986$, $p_{0.025}(\alpha_1|\b{Y})=-1.294$, $p_{0.975}(\alpha_1|\b{Y})=-0.687$), implying that women are in a worse condition with respect to men. This is in line with literature on disability, and may be attributed to the fact that women are more prone to be affected by some types of chronic diseases that are more connected with a condition of disability (e.g. they are more affected by osteoporosis and varicose veins).

As anticipated, we are mainly interested in estimating counts of people in the various disability classes for age groups within each Health district. Using MCMC output we may easily complement point estimates with measures of uncertainty. Table \ref{table:estcounts} reports the estimated counts (posterior means calculated according to (\ref{est.counts})) of people in a severe state of disability (class 5) aged 75 or more for each district of Umbria. The posterior CV and the quantiles $p_{0.025}$ and $p_{0.975}$ are the selected measures of uncertainty. The same posterior summaries are reported for the entire districts (i.e. aggregating over age groups).

\begin{figure}
\centering
\includegraphics[width=160mm]{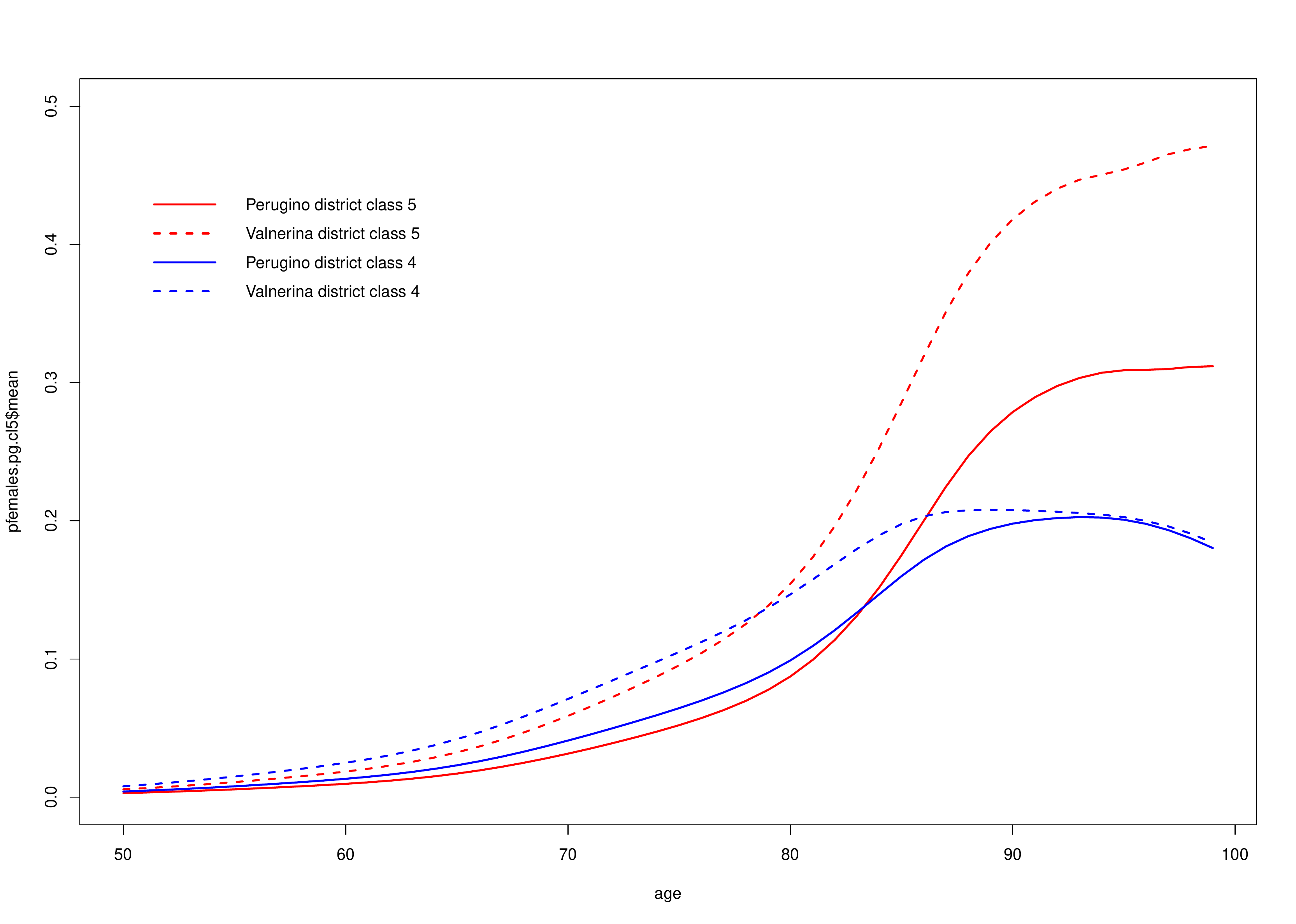}
\caption{Probability of females being classified into class 4 and class 5 for the Perugino and Valnerina districts }\label{figure:disteff}
\end{figure}

Age group specific and overall counts are not too different, as $76\%$ of people assigned to class 5 are aged 75 or more. For this reason, the posterior coefficient of variation for age-specific and district counts are close in this case. The count depends, of course, on the district population size, but we may note that for the age group $\geq 75$ the posterior mean of the probability to be in class 5 ranges from $10\%$ of the Perugino district (a mostly urban area) to the $16.7 \%$ of Val Nerina, a mountanous region. This difference partly reflects the different composition by age and gender of the elderly population in the two districts but is also influenced by estimated district effects. To evaluate the impact of the district effect, Figure \ref{figure:disteff} shows the probability of women being classified into the fourth and fifth classes versus age for the Perugino and Valnerina districts. There are large differences in these probabilities, especially for class 5, even for specific values of age. We may note that these two district are the two extreme cases: curves pertaining to others lie in between.

\begin{table}
\begin{center}
\caption{Estimation of counts of people in class 5 (complete dependency} within Health districts of Umbria with focus on the age group $\geq 75$
		\begin{tabular}[]		{l | rrrr| rrrr }
	&	 \multicolumn{ 4}{c|}{Age group $\geq 75$}& \multicolumn{ 4}{c}{All age groups}\\
District	&$\hat{Q}_{j}(5)$	&	CV	&	$p_{0.025}$	&	$p_{0.975}$	&	$\hat{Q}_{j}(5)$	&	CV	&	$p_{0.025}$	&	$p_{0.975}$\\
\hline \hline
$[11]$ Alto Tevere 	&	964.3	&	22.8	&	567	&	1431	&	1276.0	&	23.4	&	739	&	1906\\
$[12]$ Alto Chiascio	&	1108	&	20.6	&	730	&	1615	&	1454.0	&	21.5	&	946	&	2158\\
$[21]$ Perugino	&	1907	&	22.1	&	1154	&	2793	&	2528.0	&	22.4	&	1519	&	3717\\
$[22]$ Assisano	&	842.1	&	23.2	&	517	&	1285	&	1142.0	&	24.1	&	693	&	1764\\
$[23]$ Medio Tevere	&	1062	&	22.2	&	675	&	1597	&	1391.0	&	23.1	&	877	&	2131\\
$[24]$ Trasimeno	&	906	&	23.5	&	543	&	1379	&	1193.0	&	24.3	&	706	&	1848\\
$[31]$ Valnerina	&	288.1	&	27.3	&	171	&	477	&	363.6	&	29.0	&	213	&	620.7\\
$[32]$ Spoleto	&	851.7	&	20.5	&	550	&	1236	&	1106.0	&	21.1	&	709	&	1626\\
$[33]$ Foligno	&	1546	&	19.2	&	1018	&	2177	&	1985.0	&	19.5	&	1295	&	2813\\
$[41]$ Terni	&	1711	&	20.6	&	1083	&	2453	&	2242.0	&	20.7	&	1412	&	3224\\
$[42]$ Narni	&	948.7	&	21.9	&	603	&	1415	&	1256.0	&	22.7	&	790	&	1907\\
$[43]$ Orvieto	&	843.8	&	21.4	&	539	&	1246	&	1093.0	&	21.9	&	695	&	1635

\end{tabular} \label{table:estcounts}

\end{center}
\end{table}

\section{Conclusions} \label{sec:concl}

In this paper we have considered the problem of obtaining small area estimates of totals or means of a variable of interest that is not observed, but is a latent variable hidden behind a set of observed categorical (ordered) items. The motivating problem is that of estimating population counts for classes of disability from survey data. However, the proposed methodology can be applied to any other latent variable of interest. In addition, the use of latent class models allows for modeling situations in which the latent construct is unidimensional or multidimensional. 

The small area model is a unit level model defined on the probability of each unit to belong to a latent class. Multinomial logistic and proportional odds mixed models are considered and fitted as a whole together with the latent class model. One covariate is included nonparametrically in the model by using penalized splines. A hierarchical Bayesian framework is adopted to obtain estimates from the model and measures of its accuracy. 

A useful extension of p-splines in a Bayesian context is proposed that allows for eliminating the problem of posterior correlation of chains for the coefficients of the spline. In particular, we propose the use of Demmler-Reinsch basis functions and obtain the form of the variance matrix for its random coefficients. The use of this basis allows also for the inclusion of a usefully large number of knots. In fact, the number of knots when fitting p-splines with MCMC methods is usually kept very low to avoid numerical instabilities. This extension can be recommended any time p-splines are fitted using MCMC methods, also within simpler nonparametric regression models.




\bibliography{nonauto}

\begin{thebibliography}{}

\bibitem[\protect\citeauthoryear{Agresti}{Agresti}{2002}]{Agre:cate:2002}
Agresti, A. (2002).
\newblock {\em Categorical Data Analysis}.
\newblock John Wiley \& Sons.

\bibitem[\protect\citeauthoryear{Aitkin, Liu, and Chadwick}{Aitkin
  et~al.}{2009}]{Aitkin:etal:2009}
Aitkin, M., C.~Liu, and T.~Chadwick (2009).
\newblock Bayesian model comparison and model averaging for small-area
  estimation.
\newblock {\em The Annals of Applied Statistics\/}~{\em 3}, 199--221.

\bibitem[\protect\citeauthoryear{Cabrero-Garc\`ia and
  L\`opez-Pina}{Cabrero-Garc\`ia and L\`opez-Pina}{2008}]{irtadl}
Cabrero-Garc\`ia, J. and J.~A. L\`opez-Pina (2008).
\newblock Aggregated measures of functional disability in a nationally
  representative sample of disabled people: analysis of dimensionality
  according to gender and severity of disability.
\newblock {\em Quality of Life Research\/}~{\em 17}, 425--436.

\bibitem[\protect\citeauthoryear{Crainiceanu, Ruppert, and Wand}{Crainiceanu
  et~al.}{2005}]{Crai:Rupp:Wand:bugs:2005}
Crainiceanu, C., D.~Ruppert, and M.~P. Wand (2005).
\newblock Bayesian analysis for penalized spline regression using winbugs.
\newblock {\em Journal of Statistical Software\/}~{\em 14}.

\bibitem[\protect\citeauthoryear{Crespi and Boscardin}{Crespi and
  Boscardin}{2009}]{Crespi:Boscardin:2009}
Crespi, C. and W.~Boscardin (2009).
\newblock Bayesian model checking for multivariate outcome data.
\newblock {\em Computational Statistics and Data Analysis\/}~{\em 53},
  3765--3772.

\bibitem[\protect\citeauthoryear{Eilers and Marx}{Eilers and
  Marx}{1996}]{Eile:Marx:repl:1996}
Eilers, P. H.~C. and B.~D. Marx (1996).
\newblock Flexible smoothing with {B}-splines and penalties.
\newblock {\em Statistical Science\/}~{\em 11\/}(2), 89--121.

\bibitem[\protect\citeauthoryear{Erosheva}{Erosheva}{2002}]{erosh:2002}
Erosheva, E.~A. (2002).
\newblock {\em Grade of Membership and Latent Structure Models With Application
  to Disability Survey Data}.
\newblock Phd Dissertation: Department of Statistics, Carnegie Mellon
  University, Pittsburgh, PA.

\bibitem[\protect\citeauthoryear{Fabrizi and Trivisano}{Fabrizi and
  Trivisano}{2009}]{fabri:trivi:2009}
Fabrizi, E. and C.~Trivisano (2009).
\newblock Robust linear mixed models for small area estimation.
\newblock {\em Journal of Statistical Planning and Inference\/}~{\em 140},
  433--443.

\bibitem[\protect\citeauthoryear{Gelman}{Gelman}{2006}]{Gelman:2006}
Gelman, A. (2006).
\newblock Prior distributions for variance parameters in hierarchical models.
\newblock {\em Bayesian Analysis\/}~{\em 1}, 515--533.

\bibitem[\protect\citeauthoryear{Ghosh, Natarajan, Stroud, and P}{Ghosh
  et~al.}{1998}]{ghosh:jasa:98}
Ghosh, M., K.~Natarajan, T.~W.~F. Stroud, and C.~B. P (1998).
\newblock Generalized linear models for small-area estimation.
\newblock {\em Journal of the American Statistical Association\/}~{\em 93},
  273--282.

\bibitem[\protect\citeauthoryear{Jiang and Lahiri}{Jiang and
  Lahiri}{2006}]{Jian:Lahi:sae:2006}
Jiang, J. and P.~Lahiri (2006).
\newblock Mixed model prediction and small area estimation (with discussion).
\newblock {\em Test\/}~{\em 15}, 1--96.

\bibitem[\protect\citeauthoryear{Katz, Ford, Moskowitz, Jackson, and
  Jaffe}{Katz et~al.}{1963}]{Katz:1963}
Katz, S., A.~B. Ford, R.~W. Moskowitz, B.~A. Jackson, and M.~W. Jaffe (1963).
\newblock Studies of illness in the aged. the index of {ADL}: a standardized
  measure of biological and psy-chosocial function.
\newblock {\em Journal of the American Medical Association\/}~{\em 185},
  914--919.

\bibitem[\protect\citeauthoryear{Lazarsfeld and Henry}{Lazarsfeld and
  Henry}{1968}]{Laza:1968}
Lazarsfeld, P. and N.~Henry (1968).
\newblock {\em Latent structure analysis}.
\newblock Boston, Houghton Mifflin.

\bibitem[\protect\citeauthoryear{Mesbah}{Mesbah}{2004}]{mesbah:2004}
Mesbah, M. (2004).
\newblock Measurement and analysis of health related quality of life and
  environmental data.
\newblock {\em Envirometrics\/}~{\em 15}, 473--481.

\bibitem[\protect\citeauthoryear{Montanari, Ranalli, and Eusebi}{Montanari
  et~al.}{2011}]{Mont:Rana:Euse:2011}
Montanari, G.~E., M.~G. Ranalli, and P.~Eusebi (2011).
\newblock Latent variable modeling of disability in people aged 65 or more.
\newblock {\em Statistical Methods and Applications\/}~{\em 20}, 49--63.

\bibitem[\protect\citeauthoryear{Nychka and Cummins}{Nychka and
  Cummins}{1996}]{nych:cumm:96}
Nychka, D. and D.~Cummins ({1996}, {MAY}).
\newblock {Flexible smoothing with B-splines and penalties - Comment}.
\newblock {\em {STATISTICAL SCIENCE}\/}~{\em {11}\/}({2}), {104--105}.

\bibitem[\protect\citeauthoryear{Opsomer, Claeskens, Ranalli, Kauermann, and
  Breidt}{Opsomer et~al.}{2008}]{Opso:Clae:Rana:Kaue:Brei:non-:2008}
Opsomer, J.~D., G.~Claeskens, M.~G. Ranalli, G.~Kauermann, and F.~J. Breidt
  (2008).
\newblock Non-parametric small area estimation using penalized spline
  regression.
\newblock {\em Journal of the Royal Statistical Society, Series B: Statistical
  Methodology\/}~{\em 70\/}(1), 265--286.

\bibitem[\protect\citeauthoryear{Rao}{Rao}{2003}]{Rao:2003}
Rao, J. (2003).
\newblock {\em Small Area Estimation}.
\newblock Wiley Series in Survey Methodology.

\bibitem[\protect\citeauthoryear{Ruppert, Wand, and Carroll}{Ruppert
  et~al.}{2003}]{Rupp:Wan:Carr:03}
Ruppert, D., M.~P. Wand, and R.~J. Carroll (2003).
\newblock {\em Semiparametric Regression}.
\newblock Cambridge University Press, Cambridge, New York.

\bibitem[\protect\citeauthoryear{Spiegelhalter, Best, Carlin, and van~der
  Linde}{Spiegelhalter et~al.}{2002}]{Spiegel:etal:2002}
Spiegelhalter, D., N.~Best, B.~Carlin, and A.~van~der Linde (2002).
\newblock Bayesian measures of model complexity and fit (with discussion).
\newblock {\em Journal of the Royal Statistical Society, ser. B\/}~{\em 64},
  583--639.

\bibitem[\protect\citeauthoryear{Thomas, O'Hara, Ligges, and Sturz}{Thomas
  et~al.}{2006}]{Openbugs:2006}
Thomas, A., B.~O'Hara, U.~Ligges, and S.~Sturz (2006).
\newblock Making bugs open.
\newblock {\em R news\/}~{\em 6}, 12--17.

\end{thebibliography}

\end{document}